\documentclass[twocolumn]{aastex702}
\usepackage{CJK} 

\usepackage{textcomp, gensymb}
\submitjournal{AJ}

\shorttitle{H$\alpha$ in WASP-12b}
\shortauthors{Vissapragada et al.}

\begin{document}

\correspondingauthor{Shreyas~Vissapragada}
\email{svissapragada@carnegiescience.edu}

\title{WASP-12b Exhibits Persistent In-Transit H$\alpha$ Absorption}

\author[0000-0003-2527-1475]{Shreyas~Vissapragada} 
\affiliation{Carnegie Science Observatories, 813 Santa Barbara Street, Pasadena, CA 91101, USA}
\email{svissapragada@carnegiescience.edu}

\author[0009-0008-4762-6170]{Fabienne Nail}
\email{fabienne.nail@gmail.com}
\affiliation{Anton Pannekoek Institute for Astronomy, University of Amsterdam, 1090 GE Amsterdam, Netherlands}

\author[0000-0002-1417-8024]{Morgan MacLeod}
\email{morgan.macleod@cfa.harvard.edu}
\affiliation{Center for Astrophysics $\vert$ Harvard \& Smithsonian, 60 Garden Street, Cambridge, MA 02138, USA}

\author[0000-0002-3464-4993]{Adam Jensen}
\email{jensenag@unk.edu}
\affiliation{Department of Physics and Astronomy, University of Nebraska at Kearney, 2502 19th Ave, Kearney, NE 68849, USA}

\author[0000-0002-4203-4773]{Stefan Czesla}
\email{sczesla@tls-tautenburg.de}
\affiliation{Th\"{u}ringer Landessternwarte Tautenburg, Sternwarte 5, 07778 Tautenburg, Germany}

\author[0000-0003-3204-8183]{Mercedes L\'{o}pez-Morales}
\email{mlopez-morales@stsci.edu}
\affiliation{Space Telescope Science Institute, 3700 San Martin Drive, Baltimore, MD 21218, USA}

\author[0000-0002-4489-3168]{J\'{e}a Adams Redai}
\email{jea.adams@cfa.harvard.edu}
\affiliation{Center for Astrophysics $\vert$ Harvard \& Smithsonian, 60 Garden Street, Cambridge, MA 02138, USA}

\author[0000-0003-3455-8814,gname=Kevin,sname=Ortiz Ceballos]{Kevin N. Ortiz Ceballos}
\email{kortizceballos@cfa.harvard.edu}
\affiliation{Center for Astrophysics $\vert$ Harvard \& Smithsonian, 60 Garden Street, Cambridge, MA 02138, USA}

\author[0000-0001-9214-7437]{Jaehyon Rhee (\begin{CJK*}{UTF8}{mj}이재현\end{CJK*})}
\email{jrhee@cfa.harvard.edu}
\affiliation{Center for Astrophysics $\vert$ Harvard \& Smithsonian, 60 Garden Street, Cambridge, MA 02138, USA}
\affiliation{Department of Physics, Yonsei University, 50 Yonsei-ro, Seodaemun-gu, Seoul 03722, Republic of Korea}

\begin{abstract}
High-resolution H$\alpha$ and He 10830~\AA~transit spectroscopy can trace the extended upper atmospheres of exoplanets, typically revealing excess in-transit absorption that occurs in the planet's rest frame. However, hydrodynamical simulations predict that absorption should not strictly follow the planet's orbital motion if gas quickly escapes the planetary Roche lobe and shears into tidal tails. The ultra-hot Jupiter WASP-12b fills an exceptionally large fraction of its Roche lobe ($R_p/R_\mathrm{Roche} = 0.55$), and in this paper we present evidence for tidally-modulated outflow kinematics on this unique planet. We observed a transit of WASP-12b with MMT/Hectochelle and found an excess H$\alpha$ absorption of $3.29^{+0.35}_{-0.36}\%$ (9.3$\sigma$) during the transit and extending at least 30~min post-egress. The signal does not follow the planet's orbital motion and instead appears largely in the stellar rest frame, similar to the extended helium outflows in HAT-P-32b and HAT-P-67b, and is slightly redshifted ($+5.4_{-1.7}^{+1.5}$~km~s$^{-1}$). We demonstrate that a similar signal appears across previous high-sensitivity observations dating back to 2012 once the atypical signal kinematics and baseline self-subtraction are taken into account. Finally, we present a three-dimensional \textsf{Athena++} model of WASP-12b's outflowing atmosphere that simultaneously exhibits the observed extended morphology and lack of planet-aligned velocity gradient. Our study highlights the necessity of three-dimensional simulations for interpreting absorption lines formed in strong outflows, where gas may not simply follow the planet's Doppler signature.
\end{abstract}

\section{Introduction} \label{sec:intro}
The past few decades of exoplanet discovery have revealed a menagerie of close-in ($P\lesssim100$~d) planets, many of which suffer intense irradiation from their host stars. The least massive of these are highly susceptible to atmospheric photoevaporation, where high-energy radiation photoionizes and heats the upper atmosphere, driving a hydrodynamic wind that removes mass from the planetary envelope \citep[e.g.][]{Murray-Clay2009}. Due to their large gravitational potentials, hot Jupiters are typically immune to significant envelope loss through this mechanism \citep{Ionov2018, Owen2018, Vissapragada2022:helium, Lazovik2023}. However, the ultra-hot Jupiters (UHJs, here defined as Jupiter-sized planets with $P\lesssim2$~d) fill a significant fraction of their Roche lobes, and for these systems Roche-Lobe Overflow (RLO) can potentially increase the planetary mass-loss rate by orders of magnitude \citep{Valsecchi2015, Jackson2017, Koskinen2022, Thorngren2023, Hallatt2026}. The predicted mass-loss rates for RLO are exceptionally high and evolutionarily consequential, and may be responsible for setting the inner boundary of the hot Jupiter mass-period distribution \citep{Owen2018, Koskinen2022, Thorngren2023}.

Observational constraints on atmospheric loss in this extreme regime are beginning to emerge. Transit spectroscopy in the metastable helium (He$^*$) triplet at 1083~nm \citep[e.g.][]{Spake2018, Nortmann2018} and the H$\alpha$ line at 656~nm \citep[e.g.][]{Jensen2012, Yan2018} can reveal the density structure and kinematics of the upper atmosphere, providing empirical constraints on models for atmospheric loss. H$\alpha$ in particular has turned out to be remarkably productive for tracing the upper atmospheres of UHJs \citep[e.g.][]{Yan2018, Casasayas-Barris2018, Cabot2020, Borsa2021, Czesla2022}. When detected, H$\alpha$ can provide important information on the planetary mass-loss rate and outflow composition \citep{Christie2013, Wyttenbach2020, Huang2017, Yan2021, Yan2022, Huang2023}. 

H$\alpha$ and He$^*$ signals have thus far mostly been detected following the planetary radial velocity tracks, but recent 3D hydrodynamical studies have shown that this need not be the case. H$\alpha$ and He$^*$ both form at high altitudes in an outflowing atmosphere, and in exceptionally strong outflows the gas may shear into leading and lagging tails. Simulations by \citet{MacLeod2025} demonstrate that such morphologies can result in rather surprising signal kinematics. They find that the in-transit signatures of tidal streams can fall largely in the stellar rest frame, and can even adopt somewhat \textit{anti-planetary} velocity tracks (see their Figure 5). These effects have been observed: HAT-P-32b and HAT-P-67b exhibit enormous tidal tails in He$^*$, and kinematically these signatures reside close to the stellar rest frame \citep{Zhang2023, GullySantiago2024}. This is completely inconsistent with the large expected Doppler motions of these planets during transit, but in good agreement with expectations from 3D hydrodynamic simulations of these systems \citep{Zhang2023, Nail2024}. The HAT-P-32b and HAT-P-67b He$^*$ signals are also far too large in amplitude to explain away with stellar activity or the transit light source effect \citep{Cauley2017, Cauley2018, Rackham2018, Guilluy2020}, and they are clearly coincident with the optical transit times but with highly extended natures, further implicating planetary origins. Similar kinematics are possibly seen in H$\alpha$ for HAT-P-67b \citep[although it is difficult to say for sure due to limited baseline;][]{BelloArufe2023}, but not HAT-P-32b \citep{Czesla2022}, perhaps because H$\alpha$ traces a slightly lower part of the thermosphere than He$^*$ \citep[e.g.][]{Yan2022, Taylor2026}.

The inflated UHJ WASP-12b \citep{Hebb2009}, with $M_p = 1.4M_\mathrm{J}$, $R_p = 2.0R_\mathrm{J}$, $P = 1.09$~d, and $T_\mathrm{eq} = 2500$~K \citep{Wong2022}, is an interesting target for H$\alpha$ spectroscopy. It has the highest Roche-lobe filling factor of any planet yet discovered, with $f = R_p/R_\mathrm{Roche} \approx (R_p/a)(M_p/(3M_\star))^{-1/3} = 0.55$ (though WASP-19b is within 1$\sigma$), and is expected to be highly tidally deformed \citep{Li2010, Czesla2024b}, making it a particularly attractive target for observing signatures of RLO. While no evidence for H$\alpha$ was reported in the low-resolution HST transmission spectrum from \citet{Sing2013}, \citet{Jensen2018} studied the optical transmission spectrum of WASP-12b with HET/HRS, and found evidence for excess H$\alpha$ absorption during transit with an amplitude of $\sim6\%$. This remains the strongest in-transit H$\alpha$ absorption ever reported. However, HET cannot track WASP-12 across a single full transit, so the in-transit and out-of-transit spectra were collected asynchronously over the span of two months, making it more challenging to assess the timescale and velocity structure of the absorption signature. Additionally, recent observations from CAHA/CARMENES \citep{Czesla2024} and LBT/PEPSI \citep{PaiAsnodkar2024} have been unable to reproduce the HET/HRS absorption amplitude, casting further doubt on the initial signal.

Many other lines of evidence suggest an escaping atmosphere for this planet. It was one of the first planets observed to exhibit deep near-ultraviolet transits indicative of metals overflowing the planetary Roche lobe \citep{Fossati2010, Lai2010, Haswell2012, Nichols2015}, confirming predictions from models by \citet{Li2010}. The depressed Mg II line cores in the stellar spectrum may also be due to a circumstellar gas torus filled by outflowing planetary material \citep{Haswell2012, Debrecht2018}. Additionally, the double-peaked \textit{Spitzer} 4.5~$\mu$m phase curves of this planet are consistent with outflowing material emitting more strongly at 4.5~$\mu$m than 3.6~$\mu$m \citep{Cowan2012, Bell2019}. While He$^*$ was not detected in the planet's upper atmosphere with HST/WFC3 \citep[but this could be due to the low instrumental resolution;][]{Kreidberg2018} or CAHA/CARMENES \citep{Czesla2024}, it has recently been detected in Palomar/WIRC narrowband photometry \citep{Saidel2026}. 

Another important aspect of the system is that WASP-12b is on a decaying orbit, with $P/\dot{P} \approx 3$~Myr \citep{Maciejewski2016, Patra2017, Maciejewski2018, Baluev2019, Yee2020, Turner2021, Ivshina2022, Wong2022, Akinsanmi2024, Leonardi2024}. The final fate of hot Jupiters undergoing orbital decay is usually thought to be tidal disruption \citep[e.g.][]{Hamer2019}. However, RLO can fill a stellar accretion disk which back-reacts with the planet, possibly saving it from total destruction despite the significant envelope loss \citep{Valsecchi2015}. This mechanism could possibly explain the origin of extremely dense planets deep within the Neptune desert \citep{Hallatt2026}. If an H$\alpha$ signal can be detected and modeled alongside the metal absorption signals to constrain the mass-loss rate $\dot{M}_p$ \citep[as was done for WASP-121b;][]{Cabot2020, Borsa2021, Yan2021, Huang2023, Czesla2024b}, this may be the only planet for which the mass-loss timescale $M_p/\dot{M}_p$ can be compared to the orbital decay timescale $P/\dot{P}$, providing a unique insight into the ultimate fates of hot Jupiters. 

Here, we attempt to progress towards this goal by studying H$\alpha$ absorption in the atmosphere of WASP-12b using continuous timeseries high-resolution spectroscopy. By observing a full transit of WASP-12b with baseline measurements taken immediately before ingress and after egress, we attempted to constrain the timescale and velocity structure of the H$\alpha$ absorption signature. In Section~\ref{sec:obs}, we describe our transit observations of WASP-12b with the Hectochelle instrument on MMT \citep{Szentgyorgyi2011}. In Section~\ref{sec:redux}, we describe our data reduction procedures and present our final H$\alpha$ transmission spectrum. We discuss our result in the context of previous observations in Section~\ref{sec:disc}, and present a three-dimensional hydrodynamical simulation of the WASP-12b outflow that shows similar behavior to the observations in Section~\ref{sec:model}. Finally, we conclude in Section~\ref{sec:conc}.

\section{Observations} \label{sec:obs}

We observed a transit of WASP-12b on 2023 February 10 (UT) with MMT/Hectochelle. Hectochelle was a high-resolution ($R\sim38,000$), fiber-fed, multi-object spectrograph \citep{Szentgyorgyi2011} that shared an optical fiber feed and robotic positioning system with the medium-resolution Hectospec instrument \citep{Fabricant1994, Fabricant2005}. Both instruments have recently been decommissioned. Hectochelle could position up to 240 fibers ($1\farcs5$ in diameter) across a large field of view ($1\degree$ in diameter) and collected spectra in a single order simultaneously across all fibers. This setup allowed us to observe our target alongside a number of other stars spread across the field that ensured the telescope was pointed correctly. We also assigned many sky fibers across the field to ensure proper subtraction of telluric emission lines. For this observation, one fiber was assigned to the target WASP-12, 12 fibers were assigned to other stars for verifying positioning, 39 fibers were assigned to sky, and the remaining fibers were unused. For all exposures, we used a 10~min exposure time, the OB25 order-sorting filter (which is sensitive to H$\alpha$), and $2\times1$ binning, i.e. binning in the spatial direction to improve readout time and thus observational duty cycle, but not binning in the spectral direction to preserve resolution. 

After setting up the initial fiber configuration on the WASP-12 field, we proceeded to observe the star for 6.75~hr (from 02:31 UT to 09:16 UT). Using the non-linear ephemerides (i.e. accounting for the observed orbital decay) from \citet{Wong2022}, the predicted mid-transit time on this night was 05:15 UT. The prediction is precise to 0.2~min in this epoch, and agrees well with non-linear ephemerides from e.g. \citet{Yee2020}, \citet{Turner2021}, and \citet{Akinsanmi2024}. The transit duration is $2.9767\pm0.0023$~hr \citep{Turner2021}, so we covered the transit with nearly four hours of out-of-transit baseline. 

During the observation, WASP-12 passed nearly overhead at MMT. Hectochelle could not observe during this period of rapid parallactic angle change, so we paused observations for 26~min, changed the instrument rotator angle, and resumed observations with fibers reassigned. In total, we collected 36 exposures (11 prior to the rotator angle change and 25 after), for a total open-shutter time of 6~hr. Throughout the night, we also acquired ThAr arc lamp exposures for wavelength calibration. All data were processed with the standard Telescope Data Center reduction (HSRED v2.2), which included bias correction, wavelength calibration, flat fielding, telluric correction, and cosmic ray removal. 

\section{Data Analysis} \label{sec:redux}
\begin{figure}[h]
    \centering
    \includegraphics[width=\columnwidth]{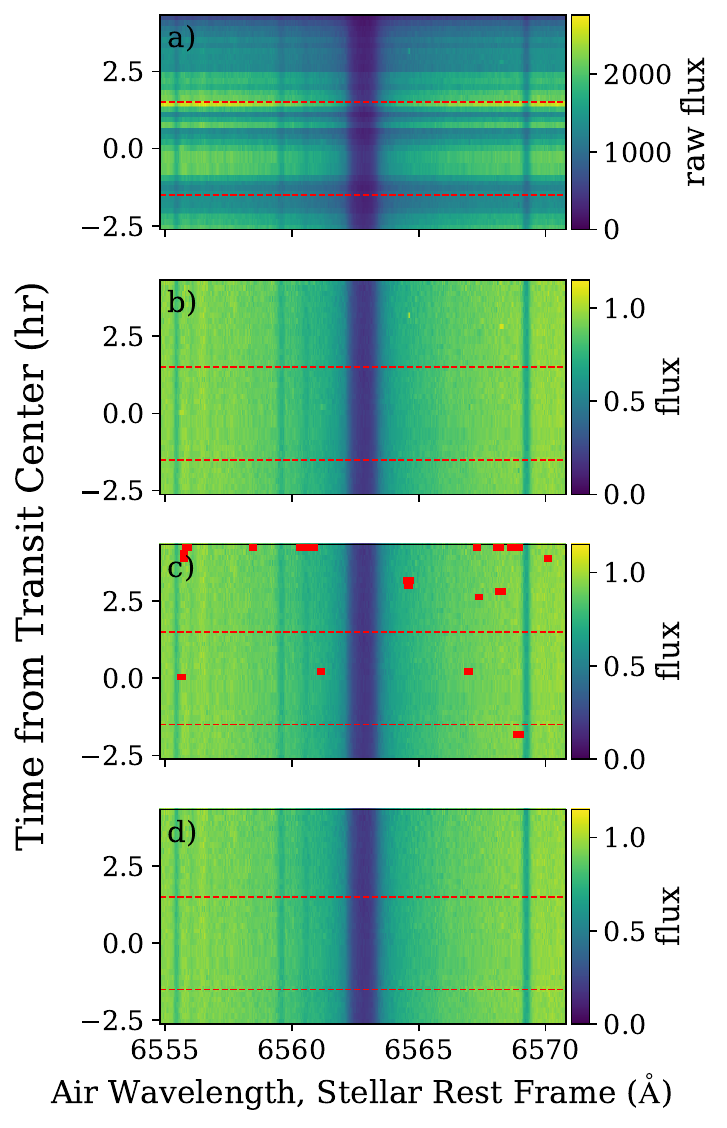}
    \caption{Timeseries spectra for WASP-12 showing the data reduction procedure. The steps shown are a) the raw data in the stellar rest frame, b) the data after continuum normalization, c) the continuum-normalized data with outliers flagged, and d) the outlier-corrected data with the fiber response variations divided out (see also Figure~\ref{fig:fibers}). The horizontal dashed lines indicate the start and end of optical transit.}
    \label{fig:reduction}
\end{figure}

The pipeline-produced spectra were in air wavelengths in the observatory rest frame. We first used \textsf{astropy} to shift the spectra into the Solar System Barycenter rest frame, and then shifted into the stellar rest frame using the systemic radial velocity of $+19.061$~km~s$^{-1}$ \citep{Husnoo2011}. We then continuum-normalized the data after clipping the lower-SNR order edges (we worked only between 6500~\AA~and 6625~\AA). We fit a fourth-order Chebyshev polynomial to each spectrum after manually masking sharp spectral lines and $\pm10$\AA~around H$\alpha$. Next, we used \textsf{spectres} \citep{Carnall2017} to resample the spectra onto an evenly-spaced wavelength grid using the median pixel spacing of $\Delta\lambda = 0.04~$\AA. We flagged remaining outliers on this grid by convolving the spectroscopic timeseries with a Laplacian of Gaussian (LoG) filter, and flagging pixels where the absolute value of the LoG image was greater than 4$\times$ the median absolute deviation of the LoG image.\footnote{This procedure is conceptually similar to \textsf{lacosmic} \citep{VanDokkum2001}, but operates on the spectral timeseries rather than in the image domain.} Outliers were corrected with bilinear interpolation (i.e. considering both the time and wavelength axes). The chain of reduction steps is shown in Figure~\ref{fig:reduction}. 

\begin{figure}[ht]
    \centering
    \includegraphics[width=\columnwidth]{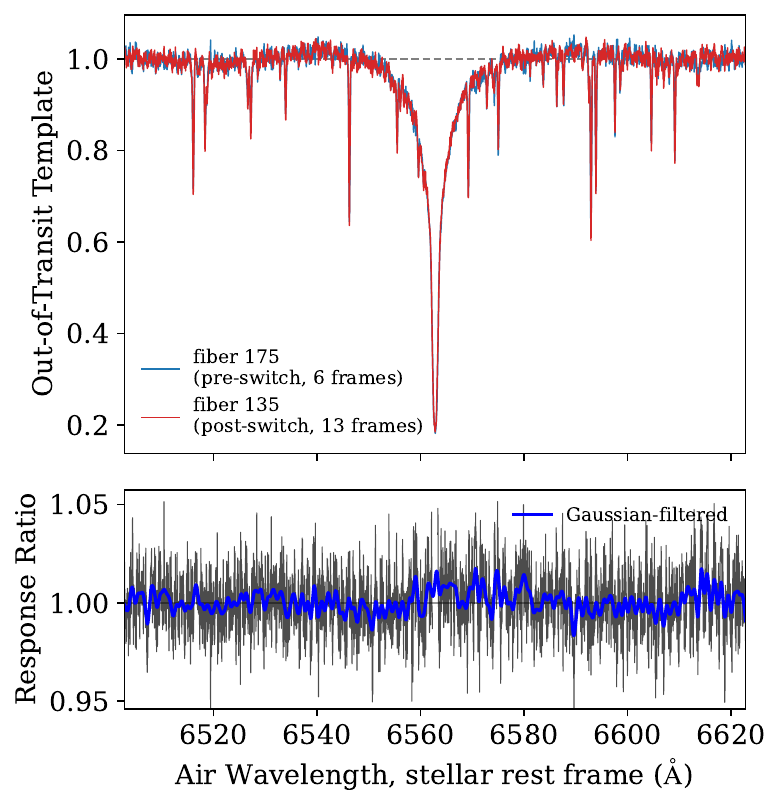}
    \caption{The top panel shows the out-of-transit templates pre- and post-fiber switch. The bottom panel shows the fiber response ratio, which is smoothed by a Gaussian filter of width $0.3$\AA~in blue.}
    \label{fig:fibers}
\end{figure}

We next separated in-transit spectra from out-of-transit spectra using the ephemerides from \citet{Wong2022}. While we initially used the optical transit duration of 2.98~hr to identify out-of-transit spectra, we saw clear evidence for strong absorption lasting 30~min after the optical transit. To avoid biasing the out-of-transit template, we only used the last two hours of data along with the pre-ingress data to construct our out-of-transit template, leaving us with 19 out-of-transit spectra. During this process, we noticed small ($\sim1\%$) systematic differences in the fiber response before and after the mid-transit fiber reassignment. We calibrated the response difference by computing the out-of-transit templates for each fiber separately and dividing them to make a noisy estimate of this fiber-flat error. We applied a Gaussian filter of width 0.3\AA~to smooth the estimate, and corrected the fiber responses by this smoothed estimate. The per-fiber out-of-transit templates and the fiber response correction are shown in Figure~\ref{fig:fibers}. 

\begin{figure}[ht!]
    \centering
    \includegraphics[width=0.5\textwidth]{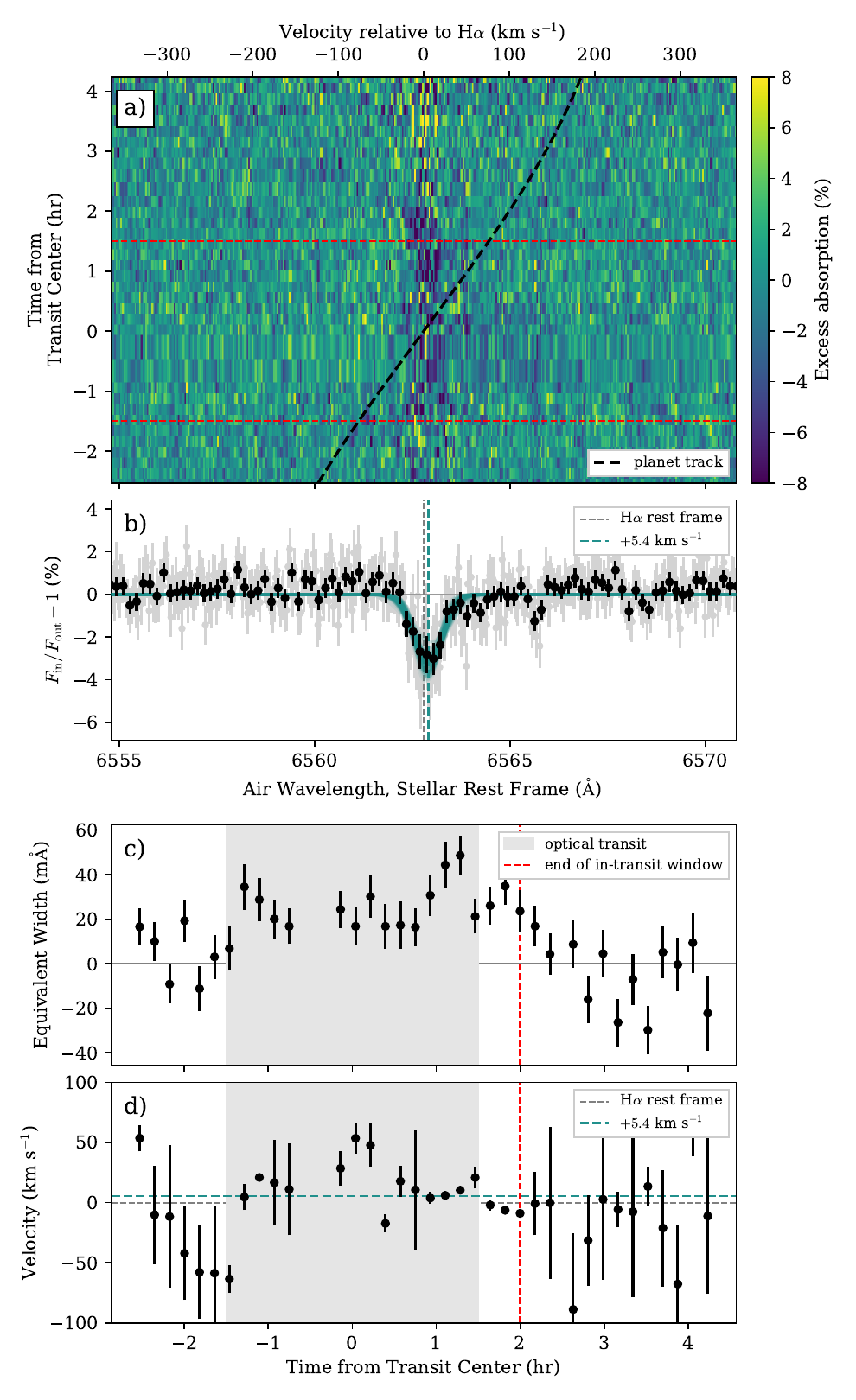}
    \caption{a) Timeseries spectroscopy of WASP-12 in the stellar rest frame. The colors indicate the difference from the average out-of-transit stellar spectrum, and red dashed lines indicate the beginning and end of the optical transit. The dashed black curve shows the H$\alpha$ line position in the planetary rest frame. b) Transmission spectrum of WASP-12b near the H$\alpha$ line (computed in the stellar rest frame). The gray points are unbinned data, the black points are binned down to the instrumental resolving power of $R\sim38,000$, and the teal curves are 100 random posterior draws from a Gaussian fit to the timeseries spectrum. The gray and teal dashed lines indicate the H$\alpha$ rest frame velocity and the median velocity from the fit, respectively. c) Equivalent width timeseries for the H$\alpha$ signal showing clear absorption during the optical transit window (gray shaded region). d) Velocity timeseries from a Gaussian fit to each excess absorption spectrum.}
    \label{fig:transmission}
\end{figure}

Finally, we divided all of the spectra in our timeseries by the combined out-of-transit template to search for planetary absorption. We show the final spectroscopic timeseries in the stellar and planetary rest frames in Figure~\ref{fig:transmission}. Excess absorption during the planetary transit is clearly visible, although it resides in the stellar rest frame rather than the planetary rest frame. To determine the statistical significance of the signal, we fit the spectroscopic timeseries treating each excess absorption spectrum as an independent measurement to which we fit a Gaussian function parameterized by an amplitude $A$ (in \%, where negative indicates absorption), a full-width at half-maximum FWHM (in km~s$^{-1}$), an overall Doppler shift $v_\mathrm{shift}$ (in km~s$^{-1}$, where negative indicates blueshift), and a free velocity gradient $\dot{v}$ (in km~s$^{-1}$~hr$^{-1}$, where negative indicates a blueshift over time). We use \textsf{emcee} \citep{ForemanMackey2013} to fit for these parameters as well as a jitter term $\sigma_\mathrm{j}$ (in \%), which is added in quadrature to the per-pixel errors. The priors and posteriors for the fit are given in Table~\ref{tab:fit}. Given that the velocity gradient was consistent with zero ($-1.7_{-1.6}^{+1.5}$~km~s$^{-1}$~hr$^{-1}$), we re-ran the fit with the velocity gradient fixed, and report both sets of posteriors. For the rest of the paper, we will adopt the fixed-gradient fit.

With a best-fitting amplitude of $A = -3.29_{-0.36}^{+0.35}$\%, we detected excess H$\alpha$ absorption during the transit at a confidence of 9.3$\sigma$. When considered alongside the continuum transit depth of 1.4\% (high-resolution observations are continuum-normalized and thus insensitive to the broadband transit), this corresponds to an equivalent opaque radius of $1.05 \pm 0.05R_\mathrm{Roche}$, suggesting ongoing RLO. We also find a significant (at about the 3$\sigma$ level) redshift of $+5.4_{-1.7}^{+1.5}$~km~s$^{-1}$. The velocity width of $38.3_{-4.3}^{+3.6}$~km~s$^{-1}$ is much larger than the instrumental resolving power ($\approx8$~km~s$^{-1}$) and similar to what has previously been seen for the UHJs KELT-9b \citep{Wyttenbach2020} and WASP-33b \citep{Yan2021}, although for these planets the H$\alpha$ absorption occurred in the planetary rather than the stellar rest frame. Finally, we attempted fitting each individual spectrum with a Gaussian function to determine if there were any kinematic trends in the data, shown in the bottom panel of Figure~\ref{fig:transmission}. The results are noisy, as one might expect given that the signal we detected averaged over the entire transit is found at only 9.3$\sigma$. Nevertheless, the signal appears to accelerate blueward during the second half of the transit and into the post-egress portion, which can be seen visually in the top panel of Figure~\ref{fig:transmission}. It is difficult to draw any further conclusions with confidence, especially because the fiber reassignment happened near the middle of the transit, which could compromise the comparison between pre- and post-reassignment velocity scales. 

\begin{deluxetable*}{lllccc}[t]
\tablewidth{0pt}
\tablecaption{H$\alpha$ Fit Parameters for WASP-12b \label{tab:fit}}
\tablehead{
\colhead{Parameter} & \colhead{Variable} & \colhead{Units} & \colhead{Prior} &
\colhead{Posterior (Gradient Free)} & \colhead{Posterior (Gradient Fixed)}}
\startdata
Amplitude    & $A$                   & \%                         & $\mathcal{U}(-10,\,10)$    & $-3.37^{+0.37}_{-0.36}$ & $-3.29^{+0.35}_{-0.36}$ \\
Full-width at half-maximum & $\mathrm{FWHM}$       & km\,s$^{-1}$               & $\mathcal{U}(0,\,100)$     & $36.8^{+3.6}_{-4.3}$    & $38.3^{+3.6}_{-4.3}$ \\
Doppler shift     & $v_\mathrm{shift}$    & km\,s$^{-1}$               & $\mathcal{U}(-50,\,50)$    & $6.5^{+1.8}_{-1.9}$     & $5.4^{+1.5}_{-1.7}$ \\
Velocity gradient & $\dot{v}$             & km\,s$^{-1}$\,hr$^{-1}$    & $\mathcal{U}(-200,\,200)$  & $-1.7^{+1.5}_{-1.6}$    & $0$ (fixed) \\
Jitter            & $\sigma_\mathrm{j}$ & \%                         & $\mathcal{U}(0,\,5)$       & $0.65\pm0.02$           & $0.65\pm0.02$\\
\enddata 
\tablecomments{Here, we summarized the posterior probability distributions using their medians and the central 68\% credible intervals.}
\end{deluxetable*}

\begin{figure*}[ht!]
    \centering
    \includegraphics[width=\textwidth]{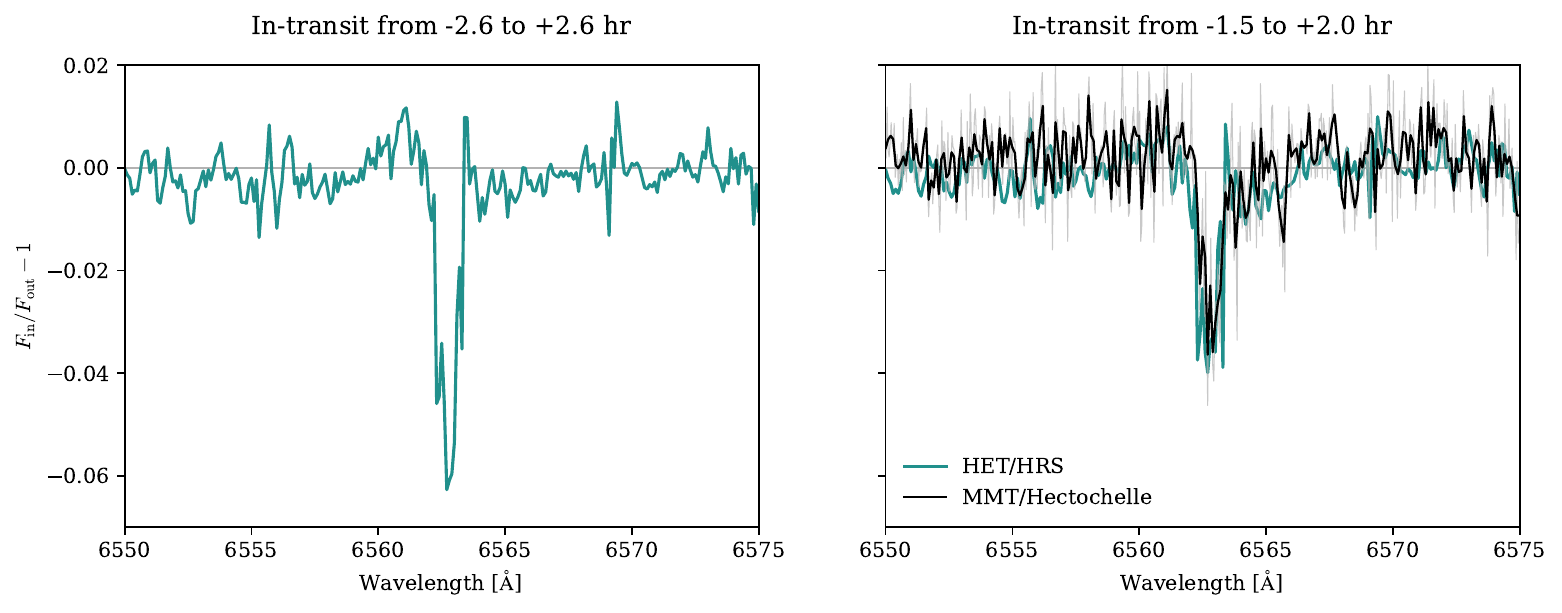}
    \caption{Comparison with the HET/HRS H$\alpha$ transit signal from \citet{Jensen2018}. On the left is our reproduction of their original result using our data reduction pipeline and their definition for ``in-transit'' exposures. On the right is the same, but now using our definition for ``in-transit'' exposures (i.e. taking self-subtraction from our shorter baseline into account) and overplotting our MMT/Hectochelle spectrum. The two spectra were taken 11 years apart and are in good agreement.}
    \label{fig:jensencomparison}
\end{figure*}

\section{Comparison with Previous Observations} \label{sec:disc}
\subsection{Agreement with \citet{Jensen2018} When Considering Self-Subtraction}
We detected excess H$\alpha$ absorption during a transit of WASP-12b, similarly to \citet{Jensen2018}. However, our $3.29\pm0.35$\% absorption signal is about half the strength of the $\sim6\%$~in-transit dip reported by this group. To ensure that the different conclusions were not due to choices in data reduction methodology, we reanalyzed the original HET/HRS spectra following our methodology described in Section~\ref{sec:redux}. Due to the lower resolving power of HET/HRS, we used a pixel spacing of $\Delta\lambda = 0.1$~\AA~when resampling the spectra, and for consistency with the original HET analysis we defined all exposures between phases of $-0.1$ and 0.1 as ``in-transit.'' Otherwise, all reduction procedures were the same. Our resulting HET/HRS transmission spectrum is shown in Figure~\ref{fig:jensencomparison}, and agrees well with Figure~9 of \citet{Jensen2018}, down to reproducing the correlated noise feature at 6570~\AA. Therefore, we can be sure that the difference between our result and the result from \citet{Jensen2018} is not due to the different data reduction strategies. \citet{PaiAsnodkar2024} performed a similar re-analysis of the \citet{Jensen2018} data and came to a similar conclusion. The HET/HRS signal is robust to choices in data reduction methodology.

An important difference between the two experiments is our definitions for in-transit and out-of-transit exposures. \citet{Jensen2018} defined in-transit exposures to be from $-2.6$~hr to $+2.6$~hr (phases of $-0.10$ to 0.10), while the in-transit absorption in our data appeared to extend from $-1.5$~hr to $+2.0$~hr (phases of $-0.06$ to $0.08$). Our MMT/Hectochelle observations only covered phases from $-0.10$ to 0.15, so it would have been challenging for us to see pre-ingress and post-egress absorption extending for the entire phase range suggested by \citet{Jensen2018}. To test whether these different definitions affected the results, we re-computed the HET/HRS transmission spectrum with our definition for in-transit exposures, which is shown on the right side of Figure~\ref{fig:jensencomparison}. Indeed, this choice significantly reduces the HET/HRS H$\alpha$ signal, and the result agrees with our measurement to within our observational errors. 

We conclude that our MMT/Hectochelle baseline is contaminated by extra pre-ingress or post-egress absorption that we were insensitive to due to limited phase coverage. This causes self-subtraction. When the self-subtraction is taken into account, we agree remarkably well with the original \citet{Jensen2018} measurement (for which the data were taken in 2012), demonstrating that the detected signal is fairly stable over a decade-long timespan. Similar situations have been reported in He~10830~\AA~observations of other hot Jupiters with high Roche-lobe-filling factors or otherwise strong outflows. The CAHA/CARMENES transits of HAT-P-32b and HAT-P-67b \citep{Czesla2022, BelloArufe2023} revealed smaller signals than HET/HPF observations that obtained more complete orbit coverage \citep{Zhang2023, GullySantiago2024}. Space-based observations are also beginning to reveal these exceptionally extended structures. \citet{Allart2025} reported helium absorption stretching for over 60\% of the orbit in WASP-121b, and \citet{Krishnamurthy2026} demonstrated that there is at least some baseline self-subtraction occurring in all existing helium observations of WASP-107b. 

\subsection{Agreement with \citet{PaiAsnodkar2024} and \citet{Czesla2024} When Considering Kinematics}

There have been two recent attempts at detecting H$\alpha$ in WASP-12b using similar methods to what we have described here. \citet{PaiAsnodkar2024} observed two transits of WASP-12b with LBT/PEPSI and \citet{Czesla2024} observed two transits of WASP-12b with CAHA/CARMENES. Both sets of transits had similar phase coverage to ours (with a few hours of baseline immediately before and after each transit measurement). Neither of these groups observed the $6\%$ signal reported by \citet{Jensen2018}. If the H$\alpha$ signal really has been fairly stable over a decade-long timespan as we have argued, then such a large signal should certainly have been seen in their measurements given the good data quality. As we have already established, part of the signal was certainly lost to baseline self-subtraction, but this is a second-order effect. At their demonstrated sensitivities, both \citet{PaiAsnodkar2024} and \citet{Czesla2024} would have readily observed a 3\% signal similar to ours.

The culprit is the rest frame. Both \citet{PaiAsnodkar2024} and \citet{Czesla2024} searched for signals in the planetary rest frame, and both groups used \textsf{SYSREM} \citep{Tamuz2005} to remove systematic artifacts due to stellar absorption and telluric contamination. Although this was a reasonable choice given that all UHJ H$\alpha$ detections thus far have followed the planetary velocity track reasonably well, we have shown that WASP-12b's H$\alpha$ transit signal morphology is highly atypical in this regard. Because \textsf{SYSREM} discriminates signals based on their velocity structure, it can efficiently remove this kinematically-atypical H$\alpha$ signal. In fact, a stellar rest-frame signal of roughly $3\%$ amplitude was removed by the \textsf{SYSREM} procedure of \citet{PaiAsnodkar2024}, as shown in the bottom-left panel of their Figure~1. Additionally, \citet{Czesla2024} observed clear H$\alpha$ variations in the stellar rest frame (see their Figure~8), which they argued could not be caused by stellar activity. The first night of their observations, in particular, reaches a depth similar to what we observe, and the temporal morphology of their signal (with a noticeable decrease in the absorption near mid-transit) matches the behavior of our signal in Figure~\ref{fig:transmission}. Their second night of observations does not agree so cleanly with ours. This may suggest that the underlying signal is variable at some level, but it is difficult to say for sure without higher-precision data. WASP-12 is a challenging target for CAHA/CARMENES transmission spectroscopy at $V = 11.6$.

\section{3D Hydrodynamical Modeling} \label{sec:model}

To aid the interpretation of the observed H$\alpha$ absorption, we performed three-dimensional hydrodynamic simulations of a planetary outflow in the WASP-12 system. Our aim is not to reproduce the observations uniquely, but rather to explore a plausible atmospheric escape scenario in which Roche-lobe overflow and orbital dynamics shape the escaping material. The simulations provide a physically-motivated example of how the outflow may be redistributed throughout the planet-star system and allow us to assess the resulting velocity field and its observational consequences.

We model the interaction between the planetary outflow, the stellar gravitational potential, and the stellar wind using the Eulerian hydrodynamics code \texttt{Athena++} \citep{Stone2020}. The simulations follow the framework presented by \citet{MacLeod2022} and \citet{Nail2024}. We adopt baseline system parameters from \citet{Leonardi2024} and \citet{Chakrabarty2019}. The computational domain is posed in spherical polar coordinates  around the star in the frame of reference that corotates with the orbit. We adopt an ideal gas equation of state with an adiabatic index of $\gamma=1.0001$, corresponding to a nearly isothermal flow. Our focus is on the dynamics of the escaping material rather than its thermochemical evolution. Consequently, the simulations do not include detailed heating, cooling, or chemical reaction networks: we note that these effects can affect the results under different assumptions. The simulations are analyzed after reaching a quasi-steady state following several planetary orbits.

The domain extends from the stellar surface at $1.69R_\odot \approx 1.175\times10^{11}$~cm to $1.175\times10^{13}$~cm or $\approx 0.79$~au and covers the full $4\pi$ solid angle. The base mesh is constructed from $144 \times 96 \times 192$ zones $(N_r, N_\theta, N_\phi)$, decomposed into $16^3$ blocks for parallelization. The zone spacing is equal in the angular coordinates and logarithmic in radius to maintain approximately cubic zone shapes.  Three additional levels of static mesh refinement (factor $8\times$ increase in linear resolution relative to the base mesh) are employed around the planet to resolve the escaping atmosphere, and one additional level is applied in a torus fully surrounding the star (factor $2\times$ increase in linear resolution). 

The calculation models interacting stellar and planetary winds, which are imposed by boundary conditions. We specify a stellar and planetary mass injection rate and use the hydrodynamic escape parameter to specify their thermal content, 
\begin{equation}
    \lambda_i = \frac{GM_i}{c_s^2 R_i},
\end{equation}
where $G$ is the gravitational constant and $c_s$ is the sound speed at the base of the outflow. The index $i$ corresponds to either the star or planet. This parameter quantifies the ratio of gravitational binding energy to thermal energy in the atmosphere, and thus determines the conditions under which a hydrodynamic wind can be launched. For the star, we adopt a nominal value of $\lambda_\ast = 15$ \citep[e.g. to describe a coronal stellar wind][]{MacLeod2022}. Because the planetary and stellar mass loss rates are not known a priori, we adopt a nominal value of $10^{13}$~g~s$^{-1}$ for the planet and $10^{12}$~g~s$^{-1}$ for the star  in this calculation. 

Motivated by the indication of a high Roche-lobe filling factor, we adopt a relatively high value of $\lambda_p = 10$, corresponding to a cool, gravitationally confined launch regime in which the planetary atmosphere is already strongly shaped by the Roche potential \citep{Nail2024}. The adopted value of $\lambda_p = 10$ corresponds to an outflow temperature of approximately $6000$\,K and a sound speed of $\approx 11$\,km\,s$^{-1}$. 

As discussed by \citet{Nail2025} and \citet{MacLeod2025}, the morphology of escaping atmospheres depends strongly on the thermodynamic state of the planetary wind. In particular, Figure~2 of \citet{Nail2025} illustrates the fundamental difference between hot and cold wind regimes. In the hot wind case, the outflow is sufficiently energetic that it expands approximately radially and remains only weakly affected by the orbital motion of the planet (creating a ``bubble"). In contrast, cold winds produce slower outflows that are strongly shaped by the planetary potential and the Roche geometry, leading to efficient redirection of the gas through the L1 and L2 Lagrange points and incorporation into Keplerian orbits. This results in extended leading and trailing structures (``streams") that can span a large fraction of the orbital path and even form complete tori \citep{MacLeod2025}.

\begin{figure}
    \centering
    \includegraphics[width=0.49\textwidth]{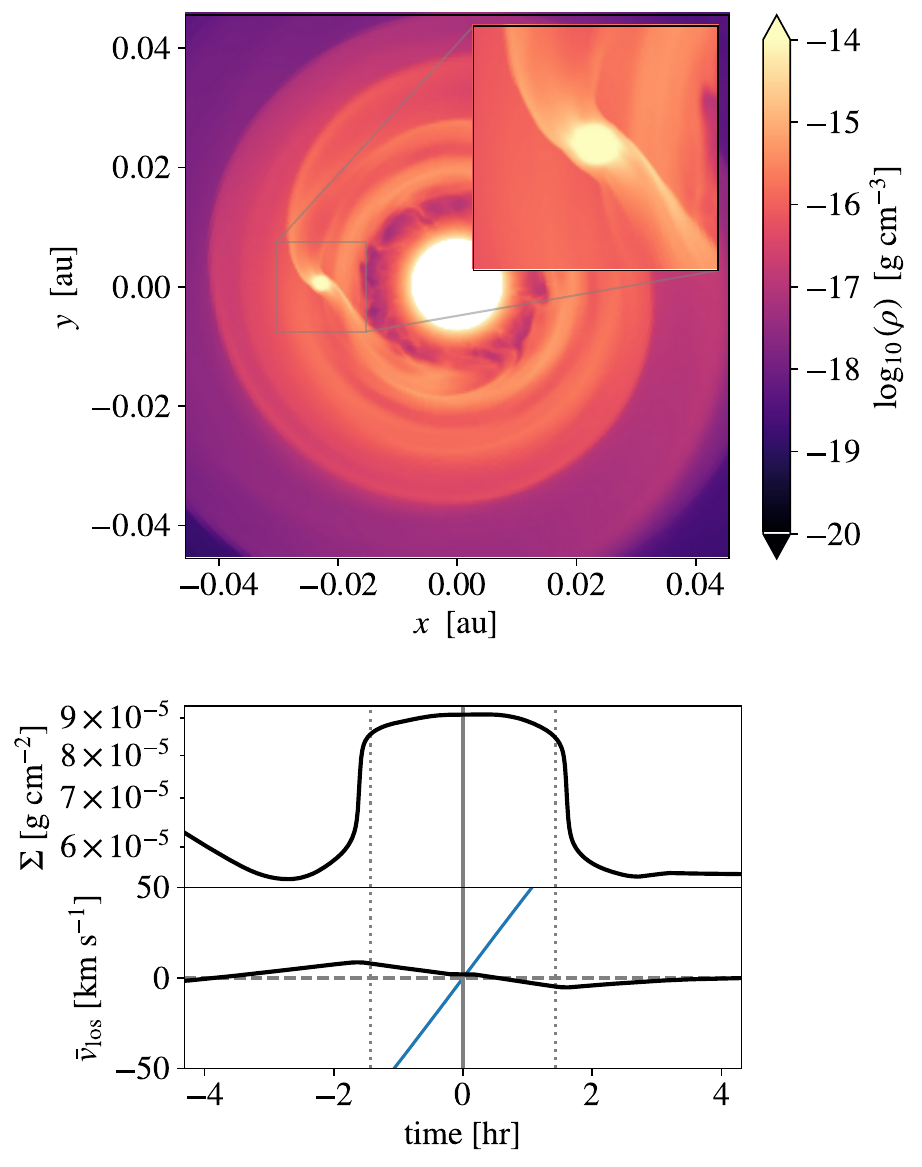}
    \caption{Hydrodynamic simulation results for the WASP-12 system. \emph{Top}: A slice through the orbital plane showing the gas density in logarithmic scale. The planetary and stellar mass loss rates are assumed to be $10^{13}$~g~s$^{-1}$ and $10^{12}$~g~s$^{-1}$, respectively, in this calculation. The host star is at the coordinate origin, and the planet is in the negative $x$-direction. Column-like tails form, both leading and trailing the planet, which then wrap into a torus surrounding the star.  \emph{Bottom}: Surface density $\Sigma$ and mass-weighted line-of-sight velocity $v_{\rm los}$ seen from an observer located in the negative $x$-direction, convolved over the stellar diameter and plotted over time relative to the transit center.  The dotted vertical lines mark the beginning and end of the optical transit, and the blue solid line indicates the planetary motion. The gas does not follow the planetary motion, but is instead dragged into a Keplerian shear of the planet-star system. A signal from this configuration is expected to align close to the stellar rest frame, which is consistent with the observational results of this study.}
    \label{fig:simulation}
\end{figure}

The upper panel of Figure~\ref{fig:simulation} shows the gas density in the orbital plane for the simulation setup described above and the adopted system parameters. Because the planetary atmosphere overfills its Roche lobe, the escaping material is not confined to a compact envelope around the planet. Instead, tidal forces and orbital motion stretch the gas into elongated streams extending both ahead of and behind the planet along its orbit.

The lower panel shows the projected surface density and line-of-sight velocity as a function of orbital phase in the orbital plane for an observer located along the negative $x$-direction. The blue line marks the projected motion of the planet, while the dotted vertical lines indicate the optical transit duration. The gas distribution clearly does not follow the planetary trajectory. After leaving the vicinity of the planet, the material is no longer gravitationally bound to the planet and becomes incorporated into the Keplerian motion of the system, spreading over a broad range of orbital phases ahead and behind the planet. Consequently, the mass-weighted line-of-sight velocity remains closer to the stellar rest frame than to the instantaneous planetary velocity.

We also observed a modest redshift in the H$\alpha$ signal during WASP-12b's transit, which could be consistent with absorption from the leading arm of the outflow (see inset). The enhanced density of this component is physically plausible, as the corresponding wind is launched from the irradiated dayside, where higher temperatures drive a stronger mass loss compared to the nightside, an effect not modeled in our hydrodynamic calculation. We note that the stellar wind can significantly affect the morphology of the escaping gas. As shown in Fig.~6 of \citet{Nail2025}, interactions between the stellar wind and the planetary outflow can confine and reshape extended streams, reducing their spatial extent and altering their density distribution. Such effects may suppress or redistribute the expected pre-transit signal, even in cases where a geometrically favorable leading-arm structure would otherwise be present.

\section{Conclusion} \label{sec:conc}
We observed a transit of WASP-12b with MMT/Hectochelle to search for planetary absorption in the H$\alpha$ line. During the transit, we found an excess absorption of $3.29_{-0.36}^{+0.35}$\% that occurred mostly in the stellar rest frame. This suggests active RLO, as the excess absorption corresponds to an equivalent opaque radius of $1.05 \pm 0.05R_\mathrm{Roche}$. The H$\alpha$ signal had a FWHM of $38.3_{-4.3}^{+3.6}$~km~s$^{-1}$ (comparable to other UHJ H$\alpha$ signal widths in the literature) and is weakly redshifted on average by $5.4_{-1.7}^{+1.5}$~km~s$^{-1}$. We see tentative evidence that the absorption signal blueshifts during the second half of the transit and into post-egress phases, but the data are not high-enough quality to say so with certainty.

Signals that appear principally in the stellar rest frame are often suspected to arise from stellar activity. However, for this H$\alpha$ signal the stability over time, coincidence with transit, highly extended nature, and large amplitude all disfavor stellar activity or the transit light source effect as possibilities \citep{Cauley2017, Cauley2018, Rackham2018, Guilluy2020}, similar to the case for HAT-P-32b and HAT-P-67b \citep{Zhang2023, GullySantiago2024, Nail2024}. After considering self-subtraction effects (i.e. the true signal duration is likely longer than our total observing baseline), we find excellent agreement between our signal and the signal originally reported by \citet{Jensen2018}, demonstrating the stability of the signal over a decade of observations with different instruments. Discrepancies between these results and the null results from \citet{PaiAsnodkar2024} and \citet{Czesla2024} are readily explained by the atypical kinematics of the signal. Both these groups searched for and ruled out H$\alpha$ signals strictly in the planetary rest frame, and there is evidence for signals of similar amplitude and spectrotemporal structure to ours in both of their datasets. Moreover, the extended nature and atypical kinematics are well-predicted by three-dimensional hydrodynamical simulations. We presented an \textsf{Athena++} model for WASP-12b's outflowing atmosphere, similar to those in \citet{MacLeod2025} and \citet{Nail2025}, that anticipates the extended outflow morphology and a signal that does not track the planetary radial velocity. 

High-resolution transmission spectroscopy pipelines are generally optimized for signals that occur in the planet rest frame. Given the proliferation of H$\alpha$ and He$^*$ detections with unexpected kinematics, we encourage investigators to also search for (and treat seriously) potential signals in or near the stellar rest frame, especially in systems where strong planetary outflows are expected. Also, the outflowing material must go somewhere, and it seems likely that strong outflows could (at least in some cases) feed broader circumstellar structures which change the apparent stellar spectrum \citep[][see also Figure~\ref{fig:simulation}]{Li2010, Fossati2010, Haswell2012, Valsecchi2015, Debrecht2018, Czesla2024, Hallatt2026, Schreyer2026}. Future searches for outflow-fed circumstellar structures --- for example, via subtle infrared excess \citep{Fossati2010b, Flagg2024} or atypical stellar spectra \citep{Haswell2012, Schreyer2026} --- would be highly valuable for clarifying the ultimate fates of planetary outflows, and perhaps even the fates of the planets themselves \citep{Hallatt2026}.

\begin{acknowledgements}
We acknowledge helpful conversations with Nelson Caldwell, ZJ Zhang, Evgenya Shkolnik, Dave Charbonneau, Vedant Chandra, Li Zeng, Seth Redfield, Antonija Oklop\v{c}i\'{c},  Morgan Saidel, Heather Knutson, and Jessica Spake. We thank Ryan Howie and Ricardo Ortiz for telescope and instrument operations and Sean Moran for developing the Hectochelle reduction code. This work made use of Claude Opus 4.8 for assistance with coding implementation. 
\end{acknowledgements}

\facilities{ADS, NASA Exoplanet Archive, MMT (Hectochelle)}
\software{\textsf{numpy} \citep{numpy},
          \textsf{scipy} \citep{scipy},
          \textsf{astropy} \citep{exoplanet:astropy13, exoplanet:astropy18},
          \textsf{matplotlib} \citep{matplotlib},
          \textsf{emcee} \citep{ForemanMackey2013},
          \textsf{spectres} \citep{Carnall2017}, \textsf{Athena++} \citep{Stone2020}}

\bibliography{references}{}
\bibliographystyle{aasjournalv7.1}

\end{document}